\documentclass
[superscriptaddress,secnumarabic,amssymb,amsmath,nobibnotes,aps,prd,showkeys,showpacs,nofootinbib]{revtex4}%
\usepackage{graphicx}
\usepackage{epsf}
\usepackage{bm}
\usepackage{amsmath}
\usepackage{amsfonts}
\usepackage{amssymb}
\usepackage{epstopdf}%
\providecommand{\U}[1]{\protect\rule{.1in}{.1in}}
\newcommand{\be}{\begin{equation}}
\newcommand{\ee}{\end{equation}}

\newcommand{\mincir}{\raise
-3.truept\hbox{\rlap{\hbox{$\sim$}}\raise4.truept\hbox{$<$}\ }}
\newcommand{\magcir}{\raise
-3.truept\hbox{\rlap{\hbox{$\sim$}}\raise4.truept\hbox{$>$}\ }}

\begin{document}
\title{Closed-form solutions of the Wheeler-DeWitt equation in a scalar-vector field
cosmological model by Lie symmetries}
\author{Andronikos Paliathanasis}
\email{anpaliat@phys.uoa.gr}
\affiliation{Instituto de Ciencias F\'{\i}sicas y Matem\'{a}ticas, Universidad Austral de
Chile, Valdivia, Chile}
\author{Babak Vakili}
\email{b.vakili@iauctb.ac.ir}
\affiliation{Department of Physics, Tehran Central Branch, IAU, Tehran, Iran}

\begin{abstract}
We apply as selection rule to determine the unknown functions of a
cosmological model the existence of Lie point symmetries for the
Wheeler-DeWitt equation of quantum gravity. Our cosmological setting consists
of a flat Friedmann-Robertson-Walker metric having the scale factor $a(t)$, a
scalar field with potential function $V(\phi)$ minimally coupled to gravity
and a vector field of its kinetic energy is coupled with the scalar field by a
coupling function $f(\phi)$. Then, the Lie symmetries of this dynamical system
are investigated by utilizing the behavior of the corresponding minisuperspace
under the infinitesimal generator of the desired symmetries. It is shown that
by applying the Lie symmetry condition the form of the coupling function and
also the scalar field potential function may be explicitly determined so that
we are able to solve the Wheeler-DeWitt equation. Finally, we show how we can
use the Lie symmetries in order to construct conservation laws and exact
solutions for the field equations.

\end{abstract}

\maketitle

\section{Introduction}

Symmetries have always played a central role in conceptual discussion of the
classical and quantum physics. The main reason may be that various laws of
conservation, such as energy, momentum, angular momentum, etc., that provide
the integrals of motion for a given dynamical system, are indeed the result of
existence of some kinds of symmetry in that system. From a more general point
of view, it can be shown that all such conservation laws are particular cases
of the so-called Noether theorem, according to which for every one-parameter
group of transformation on the configuration space of a system, which act on
the Lagrangian$~\mathcal{L}$ and preserves the Action integral, i.e., the
Euler-Lagrange equations $E_{L}\left(  \mathcal{L}\right)  $, there exists a
first integral of motion \cite{Arnold,Bluman,Stephani,Olver}. In mathematical
language this means that if the vector field $X=\xi\left(  \tau,q^{C}\right)
\partial_{\tau}+\eta^{A}\left(  \tau,q^{C}\right)  \partial_{C}$ is the
generator of the above diffeomorphism, then there exist a function $g$ such as
\cite{Bluman,Stephani}
\begin{equation}
L_{X}\mathcal{L+L}\frac{d\xi}{d\tau}=\frac{dg}{d\tau}. \label{con001}%
\end{equation}

Numerous applications of Noether's theorem\footnote{Recently in the literature
a vector field which satisfies the condition (\ref{con001}) has been termed as
\textquotedblleft Noether Gauge Symmetry\textquotedblright, see \cite{NGS}.
This is incorrect terminology, since condition (\ref{con001}) is that which
has been introduced by E. Noether in her original work. The function, $g$, of
(\ref{con001}) is a boundary term (not a gauge function) introduced to allow
for the infinitessimal transformations which in the value of the Action
Integral produced by the infinitesimal change in the boundary of the domain
caused by the infinitesimal transformation of the variables in the Action
Integral.} in general relativity and cosmology are those concerned with the
following form of action \cite{Cotsakis,Capp1,VakB,Christ,Paliat}
\begin{equation}
\mathcal{S}=\int_{\mathcal{M}}d\tau\left[  \frac{1}{2}\mathcal{G}_{AB}%
\frac{dq^{A}}{d\tau}\frac{dq^{B}}{d\tau}-\mathcal{U}(\mathbf{{q})}\right]  ,
\label{Int1}%
\end{equation}
where $q^{A}$ are the coordinates of the configuration space with metric
$\mathcal{G}_{AB}$ (the indices $A$, $B$, ... run over the dimension of this
space), $\mathcal{U}(\mathbf{{q})}$ is the potential function and $\tau$ is an
affine parameter along the evolution path of the system. In time-parameterized
theories such as general relativity, the action retains its form under time
reparameterization. Therefore, one may relate the affine parameter $\tau$ to a
time parameter $t$ by a lapse function $N(t)$ through $Ndt=d\tau$. In these
cases the action (\ref{Int1}) can be written as
\begin{equation}
\mathcal{S}=\int_{\mathcal{M}}dt\mathcal{L}(q^{A},\dot{q}^{A})=\int
_{\mathcal{M}}dtN\left[  \frac{1}{2N^{2}}\mathcal{G}_{AB}\dot{q}^{A}\dot
{q}^{B}-\mathcal{U}(\mathbf{{q})}\right]  , \label{Int2}%
\end{equation}
where an over-dot indicates derivation with respect to the time parameter $t$
and $\mathcal{L}(\mathbf{{q},\dot{{q}})}$ is the Lagrangian function of the
system. A straightforward calculation based on the Hamiltonian formalism leads
us to the Hamiltonian constraint
\begin{equation}
H=N\left[  \frac{1}{2}\mathcal{G}^{AB}P_{A}P_{B}+\mathcal{U}(\mathbf{{q}%
)}\right]  =N\mathcal{H}\equiv0, \label{Int3}%
\end{equation}
where $P_{A}$ is the momentum conjugate to $q^{A}$. Therefore, under canonical
quantization, the above constraint yields the Wheeler-DeWitt (WDW) equation
$\mathcal{H}\Psi(\mathbf{q})=0$, where $\Psi(\mathbf{q})$ is the wave function
of the quantized system and $\mathcal{H}$ should be written in a suitable
operator form. If one makes a natural choice of factor ordering, the WDW
equation may be written as%

\begin{equation}
\mathcal{H}\Psi(\mathbf{q})=\left[  \Delta+\mathcal{U}(\mathbf{q})\right]
\Psi(\mathbf{q})=0, \label{Int4}%
\end{equation}
where $\Delta=\frac{1}{\sqrt{-\mathcal{G}}}\partial_{A}\left(  \sqrt
{-\mathcal{G}}\mathcal{G}^{AB}\partial_{B}\right)  $ is the Laplacian operator
in the space with metric $\mathcal{G}_{AB}$. There is also a natural choice of
factor ordering which gives the conformally invariant WDW equation as
\cite{Wil}
\begin{equation}
\mathcal{H}\Psi=\left[  \frac{1}{2}\Delta+\frac{n-2}{8(n-1)}\mathcal{R}%
+\mathcal{U}(\mathbf{q})\right]  \Psi(\mathbf{q})=0, \label{Int5}%
\end{equation}
where $\mathcal{R}$ is the Ricci scalar obtained from the metric
$\mathcal{G}_{AB}$. In general, in terms of the $3+1$ decomposition notation,
the WDW equation which comes from the Hamiltonian constraint has the form%

\begin{equation}
\mathcal{H}\Psi=\left[  -4\kappa^{2}\mathcal{G}_{ijkl}\frac{\delta^{2}}{\delta
h_{ij}\delta h_{kl}}+\frac{\sqrt{h}}{4\kappa^{2}}\left(  -\mathcal{R}%
+2\Lambda+4\kappa^{2}T^{00}\right)  \right]  \Psi=0, \label{WDW1}%
\end{equation}
in which
\begin{equation}
\mathcal{G}_{ijkl}=\frac{1}{2\sqrt{h}}\left(  h_{ik}h_{jl}+h_{il}h_{jk}%
-h_{ij}h_{kl}\right)  , \label{WDW2}%
\end{equation}
is the the metric of superspace (the space of all 3-geometries, with metric
$h_{ij}$ and Ricci scalar $\mathcal{R}$, and the matter configuration $\Phi$).
For a scalar field, for instance, we have%

\begin{equation}
\label{WDW3}T^{00}=-\frac{1}{2h}\frac{\delta^{2}}{\delta\Phi^{2}}+\frac{1}%
{2}h^{ij}\Phi_{,i}\Phi_{,j}+V(\Phi).
\end{equation}
Note that the WDW equation (\ref{WDW1}) is not a single differential equation.
In fact, by which one has one equation in each point of the 3-dimensional
hypersurfaces. This means that (\ref{WDW1}) is in general a hyperbolic
functional differential equation on superspace. However, in the case of the
minisuperspace approximation in which we will truncate (by applying the
symmetries) the infinite degrees of freedom of the superspace to a finite
number, instead of having a WDW equation for each point of the spatial
hypersurface, we have a single WDW equation for all of them.

In this work we apply a geometric selection rule for the determination of the
potential $\mathcal{U}(\mathbf{q})$\ of \ the action integral (\ref{Int2})
which depends on the Lie symmetries of the WDW equation (\ref{Int5}). This
selection rule was proposed in \cite{AMSB} in order to find the exact solution
of a hyperbolic scalar field cosmological model in the presence of a perfect
fluid with constant equation of state parameter. In \cite{AMSB} it has been
shown that the existence of a Lie symmetry for the WDW equation is equivalent
with the existence of a conservation law for the field equations. That means
that Lie symmetries can be used in order to find invariant solutions for the
WDW equation, and study the integrability of the field equations. The plan of
the paper is as follows.

In section \ref{model} we consider a spatially flat
Friedmann--Robertson--Walker (FRW) space-time with scalar and vector fields in
a minisuperspace point of view. The minisuperspace variables turn out to
correspond to the scale factor of the universe, a scalar field minimally
coupled to gravity and a vector field coupled to scalar field by its kinetic
energy term. Such a model is studied in \cite{Maleknejad} to investigate the
possible isotropic power-law inflationary scenarios in the framework of gauge
field models. Section \ref{points} is devoted to apply the Lie symmetry method
by means of which we determine the transformations which leave the WDW
equation invariant. The application of the Lie invariants are given in section
\ref{invWDW}, where we determine all the possible solutions to the WDW
equation. In section \ref{classol}, we use the Lie symmetries of the WDW
equation to construct Noetherian conservation laws for the field equations.
Furthermore we study the integrability of the field equations and we determine
the exact cosmological solution of the model. Finally in section
\ref{conclusion} we summarize the results and draw our conclusions.

\section{The model}

\label{model}

An important issue in all of the model theories related to cosmology is to
select the matter or any extra fields used to couple with the gravitational
part of the model's action. From the standpoint of the theory of inflation,
the most widely used field has traditionally been the scalar field. Such a
field are playing an increasingly important role in the recent cosmological
models. Maybe, the mail reason is that a scalar field does not carry any
internal or external index and so, it makes it somewhat easy to work. Another
field which has seldom been studied in the literature is the vector field (or
even more general gauge fields). In general, theories studying gauge fields
coupled to gravity result in Einstein-Maxwell (or Einstein-Yang-Mills) system
which in comparison with the case of a scalar field are not easy to deal.
However, the cosmological aspects of such systems have been studied in a few
cases by a number of works, see for instance \cite{Vector} and the references
therein. Here, we pick a step further and consider a model in which in
addition of a scalar field, a vector field is also present by a special kind
of coupling with scalar field. The importance of gauge fields when is becomes
clear that we are going to explain anisotropic inflation at high energy
levels. However, as is shown in \cite{Maleknejad}, for special exponential
forms for the scalar field's potential and its coupling function with the
vector field, such models can also produce isotropic power-law inflation. By
this motivations we consider a gravity model whose dynamics is given by the
following action \cite{Maleknejad,VakVF}%
\begin{equation}
S=\int dx^{4}\sqrt{-g}\left[  \frac{1}{2}R-\frac{1}{2}g^{\mu\nu}\partial_{\mu
}\phi\partial_{\nu}\phi-V\left(  \phi\right)  -\frac{1}{4}f^{2}\left(
\phi\right)  F_{\mu\nu}F^{\mu\nu}\right]  , \label{lan.00}%
\end{equation}
where $\phi(t)$ is a homogeneous scalar field minimally coupled to gravity,
$V(\phi)$ is the potential function, and $F_{\mu\nu}$ is the strength tensor
of the vector field $A_{\mu}$ with standard definition $F_{\mu\nu}%
=2\partial_{\lbrack\mu}A_{\nu]}$. As the action shows, a coupling function
$f(\phi)$ couples the vector and scalar fields to each other. In such a
gravity model let us consider a homogeneous and isotropic cosmological model
in which the space-time is assumed to be of flat FRW whose line element can be
written
\begin{equation}
ds^{2}=-N^{2}\left(  t\right)  dt^{2}+a^{2}(t)\left(  \delta_{ij}dx^{i}%
dx^{j}\right)  , \label{lan.00a}%
\end{equation}
where $N(t)$ and $a(t)$ are the lapse function and the scale factor,
respectively. To go forward, we introduce a homogeneous vector field in the
direction of the $z$-axis: $A_{\mu}=(0;0,0,\zeta\left(  t\right)  )$, by means
of which one immediately gets $F_{\mu\nu}F^{\mu\nu}=-2N^{-2}a^{-2}\dot{\zeta
}^{2}$, see also \cite{Maleknejad}. We have assumed that the scalar field
$\phi,~$and the vector field $A_{\mu}$, inherits the symmetries of the
underlying spacetime.

With the above results at hand, the action (\ref{lan.00}) can be written in
the point-like form $S=\int dtL(\mathbf{q},\dot{\mathbf{q}})$, where
$\mathbf{q}=(a,\phi,\zeta)$ are the coordinates of the configuration space and%

\begin{equation}
\mathcal{L}\left(  a,\dot{a},\phi,\dot{\phi},\dot{\zeta}\right)  =\frac{1}%
{N}\left(  -3a\dot{a}^{2}+\frac{1}{2}a^{3}\dot{\phi}^{2}+\frac{1}{2}%
af^{2}\left(  \phi\right)  \dot{\zeta}^{2}\right)  -a^{3}NV\left(
\phi\right)  , \label{lan.01}%
\end{equation}
is an effective Lagrangian variation of which with respect to $\mathbf{q}%
^{A}=\left(  a,\phi,\zeta\right)  $, yields the corresponding Euler-Lagrange
equations for the dynamics of the model. Furthermore, in comparison with
(\ref{Int2}), from the above Lagrangian we have that the minisuperspace is,
\begin{equation}
\mathcal{G}_{AB}=diag\left(  -6a,a^{3},af^{2}\left(  \phi\right)  \right)  ,
\label{lan.01a}%
\end{equation}
the potential is,\thinspace$\mathcal{U}(\mathbf{{q})=}a^{3}V\left(
\phi\right)  $~and the Ricci scalar $\mathcal{R}$ is given from the following
expression%
\begin{equation}
\mathcal{R=}\frac{a^{-3}f^{-1}}{12}\left(  f-24f_{,\phi\phi}\right)  .
\end{equation}
By applying the standard Legendre transformation on the Lagrangian
(\ref{lan.01}) we are arrived at the following form for the Hamiltonian function%

\begin{equation}
H=N\mathcal{H}=N\left(  -\frac{p_{a}^{2}}{12a}+\frac{p_{\phi}^{2}}{2a^{3}%
}+\frac{p_{\zeta}^{2}}{2af^{2}}+a^{3}V\left(  \phi\right)  \right)  ,
\label{lan.02}%
\end{equation}
in which we have used the momenta conjugate to the dynamical variables as
\begin{equation}
p_{a}=-\frac{6}{N}a\dot{a}~,~p_{\phi}=\frac{1}{N}a^{3}\dot{\phi}~~,~p_{\zeta
}=\frac{1}{N}af^{2}\dot{\zeta}. \label{lan.03}%
\end{equation}
Now, we can construct the Hamiltonian equations of motion $\dot{\mathbf{q}%
}=\{\mathbf{q},H\}$ which are nothing but the set of equations (\ref{lan.03})
and $\mathbf{\dot{p}}_{q}=\{\mathbf{p}_{q},H\}$ which are
\begin{equation}
\dot{p}_{a}=-\frac{1}{12}\frac{p_{a}^{2}}{a^{2}}+\frac{3}{2}\frac{p_{\phi}%
^{2}}{a^{4}}+\frac{1}{2}\frac{p_{\zeta}^{2}}{a^{2}f^{2}}-3a^{2}V,
\label{lan.03a}%
\end{equation}%
\begin{equation}
\dot{p}_{\phi}=\frac{f_{,\phi}}{af^{3}}p_{\zeta}^{2}-a^{3}V_{,\phi}%
~~,~~\dot{p}_{\zeta}=0. \label{lan.03b}%
\end{equation}
The quantum version of this model may be described by the WDW equation written
in the form of relation (\ref{Int5}) which for our model takes the form
\cite{cnL}%

\begin{equation}
\left[  \Delta+\frac{n-2}{4\left(  n-1\right)  }R_{\left(  \gamma\right)
}+2Na^{3}V\left(  \phi\right)  \right]  \Psi=0, \label{lan.04}%
\end{equation}
where $\Delta=\frac{1}{\sqrt{-\gamma}}\frac{\partial}{\partial x^{A}}\left(
\sqrt{-\gamma}\gamma^{AB}\frac{\partial}{\partial x^{B}}\right)  $,
$R_{(\gamma)}$ and $n$ are: the Laplace operator, the Ricci scalar and the
dimension of the minisuperspace $\gamma_{AB}$ respectively, in which
$\gamma_{AB}$, is the conformally related with the minisuperspace
$\mathcal{G}_{AB}$, that is, $\gamma_{AB}=N^{-1}\mathcal{G}_{AB}$. Hence the
line element of $\gamma_{AB}$ is
\begin{equation}
ds_{\gamma}^{2}=N^{-1}\left(  -6ada^{2}+a^{3}d\phi^{2}+af^{2}\left(
\phi\right)  d\zeta^{2}\right)  , \label{lan.05}%
\end{equation}
where $N=$ $N=N\left(  a,\phi,\zeta\right)  $.~

The application of Noether's theorem of this model has been studied recently
in \cite{VakVF}, which some late time accelerated classical solutions are
obtained. In the following, we are going to consider the quantum aspects of
the problem by applying the Lie symmetry method on equation (\ref{lan.04}).

\section{Point symmetries and invariant functions}

\label{points}

For the convenience of the reader in the following lines we give the basic
properties and definitions of the point symmetries of differential equations
and the application of the group invariants.

Consider the partial differential equation (PDE) $\Theta=\Theta(q^{A}%
,\Psi,\Psi_{,A},\Psi_{,AB}),~\Theta=0$, which is invariant under the action of
the one parameter point transformation
\begin{equation}
\bar{q}^{A}=q^{A}+\varepsilon\xi^{A}(q^{B},\Psi)~,~~\bar{\Psi}=\Psi
+\varepsilon\eta(q^{B},\Psi),~ \label{pr.02}%
\end{equation}
that means, that there exists a function $\Lambda$ such that the following
condition holds \cite{Bluman}
\begin{equation}
\mathbf{X}^{[2]}(\Theta)=\Lambda\Theta, \label{pr.04}%
\end{equation}
or equivalently%
\begin{equation}
X^{\left[  2\right]  }\left(  \Theta\right)  =0
\end{equation}
where
\begin{equation}
\mathbf{X}^{[n]}=\mathbf{X}+\eta_{\left[  A\right]  }\partial_{\Psi_{,A}}%
+\eta_{\left[  AB\right]  }\partial_{\Psi_{AB}},
\end{equation}
is the $n$-th jet prolongation vector of the generator $\mathbf{X}$ of the one
parameter point transformation (\ref{pr.02}) \cite{Olver}. This generator
which may be written in the form
\begin{equation}
\mathbf{X}=\frac{\partial\bar{q}^{A}}{\partial\varepsilon}\partial_{q^{A}%
}+\frac{\partial\bar{\Psi}}{\partial\varepsilon}\partial_{\Psi} \label{pr.03}%
\end{equation}
and it is called, Lie point symmetry of the PDE $H~$ \cite{Bluman,Olver}. The
importance of Lie point symmetries of differential equations is that it can be
used in order to determine invariant solutions or transform solutions to
solutions (see \cite{Bluman} for details).

In what follows, we use the Lie symmetries of the WDW equation (\ref{lan.04})
as a selection rule in order to determine the unknown functions of the action
(\ref{lan.00}). Furthermore, we apply the zero order invariants in order to
find invariant solution of the WDW equation (\ref{lan.04}) and exact solution
of the field equations (\ref{lan.03})-(\ref{lan.03b}).

\subsection{Lie symmetries of the WDW equation}

\label{LsWDW}

In order to determine the Lie and the Noether symmetries of the\ WDW equation
we will follow the results of \cite{Anpaliat} which relate the point
symmetries of the WDW equation with the conformal algebra of the space which
defines the Laplace operator. This means that we can separate the problem in
two steps: (a) we will study the conformal algebra of the minisuperspace and
(b) we will determine the unknown potential.

Specifically, it has been shown that a vector field $X$ is a Lie point
symmetry for the WDW equation (\ref{lan.04}), if and only if, $X$ is a
Conformal Killing vector (CKV) of the space which defines the Laplace operator
$\Delta$, that is $X_{\left(  A;B\right)  }=\psi\left(  q^{A}\right)
\gamma_{AB}$, and \ the following condition holds%
\begin{equation}
X^{A}\left(  \mathcal{U}(\mathbf{{q})}\right)  _{;A}+2\psi\mathcal{U}%
(\mathbf{{q})}=0. \label{pr.04a}%
\end{equation}

For a general function $f\left(  \phi\right)  $ the minisuperspace
(\ref{lan.05}) admits a two dimensional conformal algebra characterized by the
vector fields $X_{1}=\partial_{\zeta}$ and $X_{H}=\frac{1}{3}a\partial
_{a}+\frac{1}{3}\zeta\partial_{\zeta}$. Hence from condition (\ref{pr.04a}) it
is easy to see that while the vector field $X_{1}$ generates a point symmetry
for the WDW equation (\ref{lan.04}) regardless of the form of the potential
$V\left(  \phi\right)  $, such a non zero potential for which $X_{H}$
generates a symmetry vector does not exist.

Now, let us to see if it is possible to choose the function $f(\phi)$ in such
a way that the minisuperspace (\ref{lan.05}) admits a greater conformal
algebra. In particular, we know that the minisuperspace (\ref{lan.05}) may has
a ten dimensional conformal algebra when it is conformally flat. On the other
hand, it is well known that the three dimensional space-times are conformally
flat if and only if the Cotton-York tensor vanishes. Therefore, by applying
this statement on (\ref{lan.05}) we are led to the following system
\begin{equation}
f_{,\phi\phi}-\frac{1}{f}f_{,\phi}^{2}=0~,~f_{,\phi\phi\phi}-\frac{f_{,\phi}%
}{f}f_{,\phi\phi}=0, \label{lan.06}%
\end{equation}
which leads to the general solution \footnote{In the following we consider
$\omega\neq0$ and $f_{0}=1.$}
\begin{equation}
f\left(  \phi\right)  =f_{0}\exp\left(  \omega\phi\right)  . \label{lan.06a1}%
\end{equation}

Furthermore, space (\ref{lan.05}) admits the same conformal algebra with
$\mathcal{G}_{AB},~$(\ref{lan.01}) which is conformal related with the
$\left(  1+2\right)  $ decomposable spacetime,%
\begin{equation}
d\bar{s}^{2}=\frac{1}{af^{2}\left(  \phi\right)  }\left(  -6ada^{2}+a^{3}%
d\phi^{2}\right)  +d\zeta^{2}. \label{lan.06a}%
\end{equation}
The last space, for arbitrary $f\left(  \phi\right)  $, admits the two CKVs
$X_{1},X_{H}$. However according to \cite{grgmt}, admits extra CKVs if and
only if the two dimensional space%
\begin{equation}
d\bar{s}_{\left(  2\right)  }^{2}=\frac{1}{af^{2}\left(  \phi\right)  }\left(
-6ada^{2}+a^{3}d\phi^{2}\right)  \label{lan.07a}%
\end{equation}
is flat, or admits gradient CKVs, the last means that (\ref{lan.07a}) is a
space of constant curvature. Hence, from the last line element we calculate
the Ricci scalar$,$ $R_{\left(  2\right)  }=2a^{-2}\left(  f_{,\phi\phi
}f-f_{,\phi}^{2}\right)  $, from where we can see that $R_{\left(  2\right)
}=0$, when $f\left(  \phi\right)  $ is given by (\ref{lan.06a1}), and there
does not exist any function $f\left(  \phi\right)  $, where $R_{\left(
2\right)  }=const\neq0$. Recall that all two dimensional spaces are Einstein
spaces and conformally flat. Hence we conclude that the WDW equation
(\ref{lan.04}) can admits Lie symmetries, if and only if, $f\left(
\phi\right)  $ is given by (\ref{lan.06a1}).

Thus, the minisuperspace takes the form
\begin{equation}
ds_{\gamma}^{2}=a^{-2\sqrt{6}\omega}\left(  -6da^{2}+a^{2}d\phi^{2}%
+e^{2\omega\phi}d\zeta^{2}\right)  , \label{lan.07}%
\end{equation}
in which without loss of generality we have chosen $N=a^{1+2\sqrt{6}\omega}$,
so that the minisuperspace becomes Ricci-flat, i.e. $R_{\left(  \gamma\right)
}=0$. Recall that the WDW equation (\ref{Int5}) is defined by the conformal
invariant Laplace operator $\hat{L}_{\gamma}=\Delta_{\gamma}+\frac
{n-2}{4\left(  n-1\right)  }R_{\gamma}$, and that under a conformal
transformation, $\bar{\gamma}_{AB}=e^{2\Omega}\gamma_{AB},$ holds, $\hat
{L}_{\bar{\gamma}}\left(  \Psi\right)  =e^{-\frac{n+2}{2}\Omega}\hat
{L}_{\gamma}\left(  e^{\frac{n-2}{2}\Omega}\Psi\right)  $. Moreover the
symmetry analysis is independent on the conformal factor $N$~$~$%
\cite{Anpaliat}.

Therefore with the use of (\ref{lan.07}) the WDW equation (\ref{lan.04}) can
be written as
\begin{equation}
\left(  -\frac{1}{6}\Psi_{,aa}+a^{-2}\Psi_{,\phi\phi}+e^{-2\omega\phi}%
\Psi_{,\zeta\zeta}\right)  +a^{-1}\left(  -\frac{\left(  1+\sqrt{6}%
\omega\right)  }{6}\Psi_{,a}+\frac{\omega}{a}\Psi_{,\phi}\right)
+2a^{4}V\left(  \phi\right)  \Psi=0. \label{lan.08}%
\end{equation}
Following \cite{Anpaliat}, we conclude that when (we consider $V_{,\phi}\neq0$)%

\begin{equation}
V\left(  \phi\right)  =V_{0}e^{-\lambda\phi} \label{lan.09}%
\end{equation}
equation (\ref{lan.08}) is invariant under the Lie algebra $G_{1}=span\left\{
X_{1},X_{2},X_{\Psi}\right\}  $ where\footnote{There is also the Lie symmetry
$X_{B}=B\left(  a,\phi,\zeta\right)  \partial_{\Psi}$, where $B\left(
a,\phi,\zeta\right)  $ is a solution of (\ref{lan.08}). However since $X_{B}$
is a trivial symmetry we will omit it.}%
\begin{equation}
X_{1}=\partial_{\zeta}~,~X_{2}=\frac{\lambda}{6}a\partial_{a}+\partial_{\phi
}+\left(  \frac{\lambda}{6}-\omega\right)  \zeta\partial_{\zeta},~X_{\Psi
}=\partial_{\Psi}. \label{lan.10}%
\end{equation}
The only non-zero commutator of the above vector fields is $\ \left[
X_{1},X_{2}\right]  =\left(  \frac{\lambda}{6}-\omega\right)  X_{1}$.

However, for special values of the constants $\omega,\lambda$ the WDW equation
(\ref{lan.08}) may admit some extra symmetries. In particular we have the
following two cases, (I): $\lambda=\omega^{-1}$,~and, (II):~$\lambda=4\omega
$,$~$with $\omega=\frac{\sqrt{6}}{3}$.

For the case (I), equation (\ref{lan.08}) is invariant under the action of the
Lie algebra $G_{2}=span\left\{  X_{1},X_{2},X_{3},X_{\Psi}\right\}  $ where
\begin{equation}
X_{3}=\frac{a\zeta}{3}\partial_{a}+2\omega\zeta\partial_{\phi}+\left(
\frac{1-6\omega^{2}}{6}\zeta^{2}+a^{2}e^{-2\omega\phi}\right)  \partial
_{\zeta}-\frac{\sqrt{6}\omega+1}{6}\zeta X_{\Psi}, \label{lan.12}%
\end{equation}
whereas for the case (II), the WDW equation is invariant under the action of
the Lie algebra $G_{3}=span\left\{  X_{1},X_{2},\bar{X}_{3},X_{4},X_{\Psi
}\right\}  ,~$in which%
\begin{equation}
\bar{X}_{3}=a\zeta e^{\frac{\sqrt{6}}{6}\phi}\left(  \frac{\sqrt{6}}%
{6}a\partial_{a}+\partial_{\phi}\right)  +\frac{\sqrt{6}}{3}a^{3}\sqrt
{6}e^{-\frac{\sqrt{6}}{2}\phi}\partial_{\zeta}%
\end{equation}
and%
\begin{equation}
X_{4}=ae^{\frac{\sqrt{6}}{6}\phi}\left(  a\partial_{a}+\partial_{\phi}\right)
.
\end{equation}

The commutators of the Lie algebras $G_{2}$, and $G_{3},$ are given in tables
\ref{comG2},~and \ref{comG3}.%

\begin{table}[tbp] \centering
\caption{The commutators of the Lie algebra $G_{2}$}%
\begin{tabular}
[c]{c|cccc}\hline\hline
$\left[  X_{A},X_{B}\right]  $ & $X_{1}$ & $X_{2}$ & $X_{3}$ & $X_{\Psi}%
$\\\hline
$X_{1}$ & $0$ & $\left(  \frac{1}{6\omega}-\omega\right)  X_{1}$ & $2\omega
X_{2}-\frac{\sqrt{6}\omega+1}{6}X_{\Psi}$ & $0$\\
$X_{2}$ & $-\left(  \frac{1}{6\omega}-\omega\right)  X_{1}$ & $0$ & $\left(
\frac{1}{6\omega}-\omega\right)  X_{3}$ & $0$\\
$X_{3}$ & $-2\omega X_{2}+\frac{\sqrt{6}\omega+1}{6}X_{\Psi}$ & $-\left(
\frac{1}{6\omega}-\omega\right)  X_{3}$ & $0$ & $0$\\
$X_{\Psi}$ & $0$ & $0$ & $0$ & $0$\\\hline\hline
\end{tabular}
\label{comG2}%
\end{table}%
%

\begin{table}[tbp] \centering
\caption{The commutators of the Lie algebra $G_{3}$}%
\begin{tabular}
[c]{c|ccccc}\hline\hline
$\left[  X_{A},X_{B}\right]  $ & $X_{1}$ & $X_{2}$ & $\bar{X}_{3}$ & $X_{4}$ &
$X_{\Psi}$\\\hline
$X_{1}$ & $0$ & $-\frac{\sqrt{6}}{9}X_{1}$ & $X_{4}$ & $0$ & $0$\\
$X_{2}$ & $-\frac{\sqrt{6}}{9}X_{1}$ & $0$ & $\frac{5\sqrt{6}}{18}X_{3}$ &
$\frac{7\sqrt{6}}{18}X_{4}$ & $0$\\
$\bar{X}_{3}$ & $-X_{4}$ & $-\frac{5\sqrt{6}}{18}X_{3}$ & $0$ & $0$ & $0$\\
$X_{4}$ & $0$ & $-\frac{7\sqrt{6}}{18}X_{4}$ & $0$ & $0$ & $0$\\
$X_{\Psi}$ & $0$ & $0$ & $0$ & $0$ & $0$\\\hline\hline
\end{tabular}
\label{comG3}%
\end{table}%

In the following, we consider the application of the Lie algebras $G_{1}$,
$G_{2},$ and $G_{3},$ in order to determine the invariant solutions of the WDW equation.

\subsection{Invariant solutions of the WDW equation}

\label{invWDW}

The WDW equation (\ref{lan.08}) is a second order PDE in which the wave
function depends on the variables $\left\{  a,\phi,\zeta\right\}  .$ This
means that in order to reduce this equation to the form of an ordinary
differential equation (ODE) we need the application of two symmetry vectors
\cite{Govinger}. To do this, we consider the invariance of the WDW equation
under the act of the: (A) Lie algebra $G_{1},$ (B) Lie algebra $G_{2}$, and
(C) Lie algebra $G_{3}$. For each case we obtain the analytical solution of
equation (\ref{lan.08}).

We would like to remark that it is possible to determine an group invariant
solution one symmetry vector, for instance see \cite{comm}. The reason for
that is that the application of a Lie symmetry in a PDE leads to a new
differential equation, which in general is independent from the original
equation, that is, new Lie symmetries can arise, a class of these symmetries
are termed as hidden symmetries, for a discussion see \cite{hid1,hid2} and
references therein.

\subsubsection{Case A: Lie invariants of $G_{1}$}

For an exponential potential function such as (\ref{lan.09}) and arbitrary
values of the constants $\lambda,\omega$, the commutation relations of the Lie
algebra $G_{1}$ allow to apply the Lie invariants of the subalgebras
$A_{\left(  i\right)  }=\left\{  X_{1},X_{2}\right\}  ~$and $A_{\left(
ii\right)  }=\left\{  X_{1},X_{2}+\beta\Psi\partial_{\Psi}\right\}
~,~\beta\in%
\mathbb{C}
$. By applying the zero order invariants of $A_{\left(  i\right)  }$ on
(\ref{lan.08}) we get the wave function as\footnote{The symbolic package Sym
for Mathematica have been used to test the results \cite{Dimas1}.}
\begin{equation}
\Psi_{A_{\left(  i\right)  }}\left(  a,\phi,\zeta\right)  =\Phi\left(
w\right)  ~~,~w=\phi-\frac{6}{\lambda}\ln a,\label{lan.13}%
\end{equation}
where $\Phi\left(  w\right)  $ satisfies the following ODE%
\begin{equation}
\left(  \lambda^{2}-6\right)  \Phi_{,ww}+\omega\lambda\left(  \sqrt{6}%
+\lambda\right)  \Phi_{,w}+2\lambda^{2}V_{0}e^{-\lambda w}\Phi
=0.\label{lan.14}%
\end{equation}

Repeating this procedure but this time with the help of the Lie invariants of
$A_{\left(  ii\right)  }$ we find the wave function%
\begin{equation}
\Psi_{A_{\left(  ii\right)  }}\left(  a,\phi,\zeta\right)  =\Phi\left(
w\right)  a^{\frac{6\beta}{\lambda}}~~,~~w=\phi-\frac{6}{\lambda}\ln a,
\label{lan.15}%
\end{equation}
where now the function $\Phi\left(  w\right)  $ should be a solution of the
following ODE
\begin{align}
0  &  =\left(  \lambda^{2}-6\right)  \Phi_{,ww}+\left(  12\beta+\omega
\lambda\left(  \sqrt{6}+\lambda\right)  \right)  \Phi_{,w}\nonumber\\
&  -\left(  6\beta^{2}+\sqrt{6}\omega\beta\lambda-2\lambda^{2}V_{0}e^{-\lambda
w}\right)  \Phi. \label{lan.16}%
\end{align}

It is seen that the solution (\ref{lan.13}) is a special case of $A_{\left(
ii\right)  }$ for $\beta=0$. Now, if we take the numerical value $~\lambda
=\pm\sqrt{6}$, the solutions of (\ref{lan.16}) can be expressed as follows
\begin{equation}
\Phi\left(  w\right)  =\Phi_{0}\exp\left(  \pm\frac{\left(  \pm3\beta
^{2}+3\omega\beta\right)  w+\sqrt{6}V_{0}e^{\mp\sqrt{6}w}}{6\left(
\omega+\beta\right)  }\right)  , \label{lan.17}%
\end{equation}
for which we are led to the following wave function as a solution of
(\ref{lan.08})
\begin{equation}
\Psi_{A_{\left(  ii\right)  }}\left(  w,\zeta\right)  =\Phi_{0}a^{\frac
{6\beta}{\lambda}}\exp\left(  \pm\frac{\left(  \pm3\beta^{2}+3\omega
\beta\right)  w+\sqrt{6}V_{0}e^{\mp\sqrt{6}w}}{6\left(  \omega+\beta\right)
}\right)  . \label{lan.18}%
\end{equation}

\paragraph{Subcase A1: $\lambda=6\omega$}

Let us now have a glance at the special case in which $\lambda=6\omega$. This
case is especially important in the sense that all of the generators of the
Lie algebra $G_{1}$ commute with each other since we get $\left[  X_{1}%
,X_{2}\right]  =0$. Therefore, we may apply on equation (\ref{lan.08}) some
additional invariants of the Lie algebras $A_{\left(  iii\right)  }=\left\{
X_{1}+\gamma\Psi\partial_{\Psi},X_{2}\right\}  $ and~$A_{\left(  iv\right)
}=\left\{  X_{1}+\gamma\Psi\partial_{\Psi},X_{2}+\beta\Psi\partial_{\Psi
}\right\}  \,.$ We continue with the application of $A_{\left(  iv\right)  }$
since when $\beta=0$, $A_{\left(  iii\right)  }=A_{\left(  iv\right)  }$. With
the same steps as in the previous section we will have%

\begin{equation}
\Psi_{A_{\left(  iv\right)  }}\left(  a,\phi,\zeta\right)  =\Phi\left(
w\right)  a^{\frac{\beta}{\omega}}e^{\gamma\zeta}~,~w=\phi\mp\sqrt{6}\ln a,
\label{lan.19}%
\end{equation}
in which
\begin{align}
0  &  =\left(  6\omega^{2}-1\right)  \Phi_{,ww}+\left(  2\beta+\sqrt{6}%
\omega^{2}\left(  1+\sqrt{6}\omega\right)  \right)  \Phi_{,w}\nonumber\\
&  -\left(  \beta^{2}+\sqrt{6}\beta\omega^{2}-6\gamma^{2}\omega^{2}e^{-2\omega
w}-12\omega^{2}V_{0}e^{-6\omega w}\right)  \Phi. \label{lan.20}%
\end{align}
If we take $\lambda=\pm\sqrt{6}$, i.e. $\omega^{2}=\frac{1}{6}$, this equation
admits an exact solution as
\begin{equation}
\Phi^{+}\left(  w\right)  =\Phi_{0}\exp\left(  \frac{\left(  \sqrt{6}%
\beta+6\beta^{2}\right)  w+3\sqrt{6}\gamma^{2}e^{-\frac{\sqrt{6}}{3}w}%
+2\sqrt{6}V_{0}e^{-\sqrt{6}w}}{2\left(  \sqrt{6}+6\beta\right)  }\right)
,~\lambda=\sqrt{6}, \label{lan.21}%
\end{equation}%
\begin{equation}
\Phi^{-}\left(  w\right)  =\Phi_{0}\exp\left(  \frac{\left(  \sqrt{6}%
\beta+6\beta^{2}\right)  w-3\sqrt{6}\gamma^{2}e^{\frac{\sqrt{6}}{3}w}%
-2\sqrt{6}V_{0}e^{\sqrt{6}w}}{12\beta}\right)  ~,~\lambda=-\sqrt{6}.
\end{equation}
We note that when $\gamma=0$, the solution (\ref{lan.19}) is the invariant
solution of the Lie algebra $A_{\left(  iii\right)  }$.

\subsubsection{Case B: Lie invariants of $G_{2}$}

In this section we consider again the exponential potential (\ref{lan.09}) but
now we assume the relation $\lambda=\frac{1}{\omega}$ between the constants
$\lambda$ and $\omega$. In this case the WDW equation (\ref{lan.08}) is
invariant under the action of the Lie algebra $G_{2}$. From table \ref{comG2},
we see that there exist the following two dimensional solvable subalgebras:
$A_{\left(  i\right)  },~A_{\left(  ii\right)  },~B_{\left(  i\right)
}=\left\{  X_{2},X_{3}\right\}  ~$and $~B_{\left(  ii\right)  }=\left\{
X_{2}+\beta X_{\Psi},X_{3}\right\}  $ for arbitrary values of $\lambda$ and
when $\lambda^{2}=6$ the extra subalgebras $A_{\left(  iii\right)
},~A_{\left(  vi\right)  },$ $B_{\left(  iii\right)  }=\left\{  X_{2}%
,X_{3}+\delta X_{\Psi}\right\}  $ and $B_{\left(  iv\right)  }=\left\{
X_{2}+\beta X_{\Psi},X_{3}+\delta X_{\Psi}\right\}  $ will be added to the
above list.

For arbitrary constant $\lambda$, the application of $A_{\left(  i\right)  }$
and $A_{\left(  ii\right)  }$ is studied before, so we will investigate the
reduction with the more general subalgebra $B_{\left(  ii\right)  }$
application of which on (\ref{lan.08}) results
\begin{equation}
\Psi_{B_{\left(  ii\right)  }}\left(  a,\phi,\zeta\right)  =a^{6\beta
\omega-2\rho}\left(  6a^{2}+\left(  6\omega^{2}-1\right)  e^{2\omega\phi}%
\zeta^{2}\right)  \Phi\left(  w\right)  ~,~w=\phi-6\omega\ln a, \label{lan.23}%
\end{equation}
where $\rho=\frac{12\beta\omega+\sqrt{6}\omega+1}{2\left(  1-6\omega
^{2}\right)  }$, $~\omega^{2}\neq\frac{1}{6}~$ and $\Phi\left(  w\right)  $
satisfies the second order ODE
\begin{align}
0  &  =\left(  1-6\omega^{2}\right)  \Phi_{,ww}+6\omega\left(  12\beta
\omega+\sqrt{6}\omega+1\right)  \Phi_{,w}+\nonumber\\
&  +\left(  12V_{0}-\left(  1+6\beta\omega\right)  \left(  1+6\beta
\omega+\sqrt{6}\omega\right)  e^{\frac{w}{\omega}}\right)  \Phi.
\label{lan.24}%
\end{align}
In what follows some solutions are presented for the special value
$\lambda^{2}=6$.

\paragraph{Subcase B1: $\ \lambda^{2}=6$}

As we have mentioned above when $\lambda=\pm\sqrt{6}$, i.e. $\omega=\pm
\frac{\sqrt{6}}{6},$ we may study the reduction of equation (\ref{lan.08}) by
using the Lie subalgebra $B_{\left(  iv\right)  }$. Thus, we apply $B_{\left(
iv\right)  }$ on (\ref{lan.08}) to obtain the wave functions as:%

\begin{equation}
\Psi_{+B_{\left(  iv\right)  }}=a^{\sqrt{6}\beta}\exp\left(  -\frac{\left(
\sqrt{6}\beta\zeta+\zeta-6\delta\right)  ze^{\frac{\sqrt{6}}{3}\phi}}{6a^{2}%
}\right)  \Phi_{+}\left(  w_{-}\right)  ~,~~w_{-}=\phi-\sqrt{6}\ln a,
\label{lan.25}%
\end{equation}
with
\begin{equation}
\Phi_{+}\left(  w_{-}\right)  =\Phi_{0+}\exp\left[  \frac{\sqrt{6}}{2\left(
\sqrt{6}+6\beta\right)  }\left(  2V_{0}e^{-\sqrt{6}w_{-}}-3\delta^{2}%
e^{\frac{\sqrt{6}}{3}w_{-}}\right)  +\frac{3\beta+\sqrt{6}}{6}w_{-}\right]  ,
\label{lan.26}%
\end{equation}
for $\lambda=\sqrt{6}$ and%

\begin{equation}
\Psi_{-B_{\left(  iv\right)  }}=a^{-\sqrt{6}\beta}\exp\left(  \frac{\left(
\sqrt{6}\beta\zeta+6\delta\right)  e^{-\frac{\sqrt{6}}{3}\phi}}{6a^{2}%
}\right)  \Phi_{-}\left(  w_{+}\right)  ~,~w_{+}=\phi+\sqrt{6}\ln a,
\label{lan.27}%
\end{equation}
with
\begin{equation}
\Phi_{-}\left(  w_{+}\right)  =\Phi_{0-}\exp\left[  -\frac{\sqrt{6}}{12\beta
}\left(  2V_{0}e^{\sqrt{6}w_{+}}-3\delta^{2}e^{\frac{\sqrt{6}}{3}w_{+}%
}\right)  +\frac{\left(  6\beta-\sqrt{6}\right)  }{12}w_{+}\right]  ,
\label{lan.28}%
\end{equation}
and for $\lambda=-\sqrt{6}$. Finally we would like to emphasize that since the
WDW equation is linear, the general invariant solution with the potential
(\ref{lan.09}) with $\lambda^{2}=6$~and$~\omega=\frac{1}{\lambda}$ is
\begin{equation}
\Psi_{\Sigma}\left(  a,\phi,\zeta\right)  =\sum_{\bar{\beta},\bar{\gamma}}%
\bar{\Psi}_{\pm A_{\left(  iv\right)  }}+\sum_{\beta,\delta}\Psi_{\pm
B_{\left(  iv\right)  }}, \label{lan.30}%
\end{equation}
where $\bar{\Psi}_{\pm A_{\left(  iv\right)  }}$ are the invariant solution
(\ref{lan.19}) with respect to the subalgebra $A_{\left(  iv\right)  }$.

\subsubsection{Case C: Lie invariants of $G_{3}$}

Consider now the special case where $\lambda=4\omega,$ in the exponential
potential (\ref{lan.09}), and $\omega=\frac{\sqrt{6}}{3}$. In that case the
WDW equation (\ref{lan.08}) is invariant under the group of transformation
which form the Lie algebra $G_{3}$. We apply the extra subalgebras $C_{\left(
i\right)  }=\left\{  X_{1}+\gamma X_{\Psi},X_{4}+\delta X_{\Psi}\right\}
$,$~C_{\left(  ii\right)  }=\left\{  X_{2}+\beta X_{\Psi},X_{3}\right\}
$,~$C_{\left(  iii\right)  }=\left\{  X_{3},X_{4}+\rho X_{\Psi}\right\}
$,~$C_{\left(  iv\right)  }=\left\{  X_{2}+\beta X_{\Psi},X_{3}+cX_{4}%
\right\}  $

From the Lie algebra $C_{\left(  i\right)  }$ we find that,
\begin{equation}
\Psi_{C_{\left(  i\right)  }}\left(  a,\phi,\zeta\right)  =e^{-\frac{\sqrt{6}%
}{6}\phi}e^{\gamma\zeta}\exp\left(  \frac{\sqrt{6}\gamma^{2}}{6\delta}%
a^{3}e^{-\frac{\sqrt{6}}{2}\phi}-\frac{\sqrt{6}\delta}{2a}e^{-\frac{\sqrt{6}%
}{6}\phi}+\frac{a^{7}}{14\delta}e^{-\frac{7\sqrt{6}}{6}\phi}\right)  ,
\end{equation}
while the Lie algebra $C_{\left(  ii\right)  }$ give us the solution%
\begin{equation}
\Psi_{C_{\left(  ii\right)  }}=\Phi\left(  u\right)  e^{-3\beta\phi}%
a^{3\sqrt{6}\beta},
\end{equation}
where $\Phi\left(  u\right)  $ is given by the following equation%
\begin{equation}
u\Phi_{,uu}+\left(  10+\frac{3\sqrt{6}}{7}\beta\right)  \Phi_{,u}+\frac{3}%
{28}\Phi=0
\end{equation}
and $u=a^{4}\left(  2a^{2}e^{\frac{2}{3}\sqrt{6}\phi}+\zeta^{2}e^{-\frac{2}%
{3}\sqrt{6}\phi}\right)  $.

Furthermore, from the application of $C_{\left(  iii\right)  }$, we find
\begin{equation}
\Psi_{C_{\left(  iii\right)  }}=\exp\left(  \frac{\sqrt{6}}{84\beta}\left(
6e^{-\frac{7\sqrt{6}}{6}v}+7\beta v\right)  \right)  \exp\left(  -\frac
{\sqrt{6}\beta}{4a^{3}}z\right)
\end{equation}
where $z=\left(  2a^{2}e^{-\frac{\sqrt{6}}{6}\phi}+\zeta e^{\frac{\sqrt{6}}%
{2}\phi}\right)  $, and $v=\phi-\sqrt{6}\ln a$. From the rest of the Lie
subalgebras of $G_{5}$, we derive trivial solutions for the wavefunction
\ $\Psi$.

As we discussed above, it is possible to determine invariant solution by
applying the zero-order invariants of only one of the Lie symmetry vectors.
For example, the application of the Lie symmetry vector $X_{4}$, in
(\ref{lan.08}), reduce the WDW equation to the following second order
differential equation%
\begin{equation}
\Phi_{,\zeta\zeta}+6\Phi e^{-\frac{2\sqrt{3}}{3}v}=0,
\end{equation}
where $\Psi\left(  a,\phi,\zeta\right)  =\Phi\left(  v,\zeta\right)  $,
and$~v=\phi-\sqrt{6}\ln a\,.$ From the last we find the solution%
\begin{equation}
\Phi\left(  v,\zeta\right)  =\Phi_{1}\left(  v\right)  \sin\left(
e^{-\frac{2\sqrt{3}}{3}v}\right)  +\Phi_{2}\left(  v\right)  \sin\left(
e^{-\frac{2\sqrt{3}}{3}v}\right)  .
\end{equation}

Below we continue with the determination of the analytical solution of the
field equations for that model.

\section{Conservation laws and analytical solutions of the field equations}

\label{classol}

In \cite{Anpaliat}, it has been shown how to extract the conservation laws and
compute the Noether integrals for a given classical Lagrangian from the Lie
symmetries of the WDW equation. In this section, for each Lie group of
transformations in terms of which the WDW equation (\ref{lan.08}) is
invariant, we are going to find the conservation laws correspond to the
classical Lagrangian (\ref{lan.01}) and the corresponding lapse functions
$N\left(  a,\phi,\zeta\right)  $.

To start let us first consider the Lie algebra $G_{1}$. In this case
Lagrangian (\ref{lan.01}) has the following integrals
\begin{equation}
I_{1}=\frac{1}{N}ae^{2\omega\phi}\dot{\zeta},~~\text{with }N=N_{1}\left(
a,\phi,\zeta\right)  ,~\text{ } \label{lan.31}%
\end{equation}%
\begin{equation}
I_{2}=\frac{1}{N}\left(  \lambda a^{2}\dot{a}+a^{3}\dot{\phi}+a\zeta
e^{2\omega\phi}\left(  \frac{\lambda}{6}-\omega\right)  \dot{\zeta}\right)
,~\text{with}~N=N_{2}\left(  a,\phi,\zeta\right)  , \label{lan.32}%
\end{equation}
where $N_{1}\left(  a,\phi,\zeta\right)  =N_{1}\left(  a,\phi\right)
e^{-2\mu\zeta}$ and $N_{2}\left(  a,\phi,\zeta\right)  =N_{2}\left(
\phi-\frac{6}{\lambda}\ln a,\zeta a^{-\frac{\lambda-6\omega}{\lambda}}\right)
a^{3-\frac{12\mu}{\lambda}},$ and $\mu\in%
\mathbb{R}
$. However, when $\lambda=\frac{1}{\omega}$, from the Lie algebra $G_{2}$ we
have an extra conservation law which comes from the Lie symmetry $X_{3}$, that
is
\begin{equation}
I_{3}=\frac{1}{N}\left(  -2a^{2}\zeta\dot{a}+2\omega a^{3}\zeta\dot{\phi
}+\left(  \frac{\left(  1-6\omega^{2}\right)  }{6}ae^{2\omega\phi}%
+a^{3}\right)  \dot{\zeta}\right)  ,~\text{with }N=N_{3}\left(  a,\phi
,\zeta\right)  , \label{lan.33}%
\end{equation}
where $N_{3}\left(  a,\phi,\zeta\right)  =N\left(  \phi-6\omega\ln
a,\frac{e^{-2\omega\phi}}{6\omega^{2}-1}a^{6\omega^{2}-1}\left[  \left(
6\omega^{2}-1\right)  \zeta^{2}e^{2\omega\phi}+6a^{2}\right]  \right)  $.

It should be noted that the integrals $\left\{  I_{1},I_{2},I_{3}\right\}  $
satisfy a Poisson algebra which has the same structure as the commutative
algebra of the symmetry vectors $\left\{  X_{1},X_{2},X_{3}\right\}  $. This
means that the dynamical system which is described by Lagrangian
(\ref{lan.01}) admits three independent conservation laws. On the other hand,
since (\ref{lan.01}) is "time-independent", the Hamiltonian constraint
(\ref{lan.02}), i.e. $H=h=0$, shows itself also as a conservation law which
here is nothing but the first Friedmann equation (\ref{lan.02}). It is easy to
see that when $\lambda=6\omega,~$the dynamical system with Lagrangian
(\ref{lan.01}) is Liouville integrable with respect to the integrals $\left\{
h,I_{1},I_{2}\right\}  $. Moreover, when $\omega^{2}=\frac{1}{6}$, this
Lagrangian admits the Noether integrals $\left\{  h,I_{1},I_{2},I_{3}\right\}
$. Finally when $\lambda=4\omega$, and $\omega=\frac{\sqrt{6}}{6}$, then from
the elements of the Lie algebra $G_{3}$, new Noetherian conservation laws
arise, it is easy to see that in that case the field equations form a
Liouville integrable system.

Below we determine the analytical solution of the field equations for the case
where $\lambda=6\omega$.

\subsection{Analytical solution of the field equations for $\lambda=6\omega$}

In order to determine the exact solution of the field equations we will use
the Hamilton-Jacobi method to reduce the Hamiltonian system. For the
Hamiltonian (\ref{lan.02}) with the potential function (\ref{lan.09}) if we
choose $\lambda=6\omega$, the (null) Hamilton-Jacobi equation takes the form
\begin{equation}
-\frac{1}{12a}\left(  \frac{\partial S}{\partial a}\right)  ^{2}+\frac
{1}{2a^{3}}\left(  \frac{\partial S}{\partial\phi}\right)  ^{2}+\frac
{1}{2ae^{2\omega\phi}}\left(  \frac{\partial S}{\partial\zeta}\right)
^{2}+V_{0}a^{3}e^{-6\omega\phi}=0, \label{lan.34}%
\end{equation}
where $S=\left(  a,\phi,\zeta\right)  $ and $p_{a}=\frac{\partial S}{\partial
a}$, $p_{\phi}=\frac{\partial S}{\partial\phi}$,~$p_{\zeta}=\frac{\partial
S}{\partial\zeta}$. To go forward, let us introduce the following coordinate
transformations
\begin{equation}
a=e^{\omega x},~~\phi=x+y~,~\zeta=\zeta. \label{lan.35}%
\end{equation}
In terms of these new variables the Hamiltonian (\ref{lan.02}) becomes
\begin{equation}
0=-\frac{e^{-3\omega x}}{6\omega^{2}}\left[  p_{x}^{2}-2p_{x}p_{y}+\left(
1-6\omega^{2}\right)  p_{y}^{2}-e^{-2y}p_{\zeta}^{2}\right]  +2V_{0}%
e^{\omega\left(  3x-6y\right)  }. \label{lan.35a}%
\end{equation}
Also, the Noether conservation laws (\ref{lan.31}) and (\ref{lan.32}) are
\begin{equation}
I_{1}=p_{\zeta}~,~I_{2}=p_{x}. \label{lan.36}%
\end{equation}
Now, one is led to the following field equations
\begin{equation}
\dot{x}=-\frac{1}{6}\left(  Ne^{3\omega x}\right)  ^{-1}\omega^{2}\left(
p_{x}-p_{y}\right)  , \label{lan..36a}%
\end{equation}%
\begin{equation}
\dot{y}=\frac{1}{6}\left(  Ne^{3\omega x}\right)  ^{-1}\left(  p_{x}+\left(
6\omega^{2}-1\right)  p_{y}\right)  , \label{lan..36b}%
\end{equation}%
\begin{equation}
\dot{\zeta}=p_{\zeta}~\left(  Ne^{3\omega x+2\omega y}\right)  ^{-1}.
\label{lan..36c}%
\end{equation}
Furthermore, the Hamilton-Jacobi equation (\ref{lan.34}) in the new
coordinates (\ref{lan.35}) takes the form
\begin{equation}
0=-\frac{e^{-3\omega x}}{6\omega^{2}}\left[  \left(  \frac{\partial
S}{\partial x}\right)  ^{2}-2\left(  \frac{\partial S}{\partial x}\right)
\left(  \frac{\partial S}{\partial y}\right)  +\left(  1-6\omega^{2}\right)
\left(  \frac{\partial S}{\partial y}\right)  ^{2}-e^{-2\omega y}\left(
\frac{\partial S}{\partial z}\right)  ^{2}\right]  +2V_{0}e^{\omega\left(
3x-6y\right)  }. \nonumber\label{lan.37}%
\end{equation}
By using the integrals (\ref{lan.36}), we may separate the action function
$S\left(  x,y,\zeta\right)  $ as $S\left(  x,y,\zeta\right)  =S_{1}\left(
x\right)  +S_{2}\left(  y\right)  +S_{3}\left(  \zeta\right)  ,$ where
\begin{equation}
S_{1}\left(  x\right)  =I_{2}x+x_{0},~S_{3}\left(  \zeta\right)  =I_{1}%
\zeta+\zeta_{0},
\end{equation}
and
\begin{equation}
S_{2}\left(  y\right)  =-\frac{I_{2}y}{6\omega^{2}-1}-\left(  \frac{\sqrt{6}%
}{6\omega^{2}-1}\right)  \int\sqrt{\omega^{2}I_{2}^{2}-I_{1}\omega^{2}\left(
6\omega^{2}-1\right)  e^{-2\omega y}-2V_{0}\omega^{2}\left(  6\omega
^{2}-1\right)  e^{-6\omega y}}dy, \label{lan.38}%
\end{equation}
for $\omega^{2}\neq\frac{1}{6}$ and
\begin{equation}
S_{2}\left(  y\right)  =\frac{I_{2}}{2}y\pm\frac{V_{0}}{6I_{2}}e^{-\sqrt{6}%
y}\pm\frac{\sqrt{6}I_{1}^{2}}{4I_{2}}e^{-\frac{\sqrt{6}}{3}y}, \label{lan.38a}%
\end{equation}
for $\omega^{2}=\frac{1}{6}.$ Therefore, the field equations will be reduced
to the following system
\begin{equation}
\dot{x}=-\frac{1}{6}\left(  Ne^{3\omega x}\right)  ^{-1}\omega^{2}\left(
I_{2}-\left(  \frac{\partial S}{\partial y}\right)  \right)  , \label{lan.39}%
\end{equation}%
\begin{equation}
\dot{y}=\frac{1}{6}\left(  Ne^{3\omega x}\right)  ^{-1}\left(  I_{2}+\left(
6\omega^{2}-1\right)  \left(  \frac{\partial S}{\partial y}\right)  \right)  ,
\label{lan.40}%
\end{equation}%
\begin{equation}
\dot{\zeta}=I_{1}\left(  Ne^{3\omega x}\right)  ^{-1}e^{-2\omega y}.
\label{lan.41}%
\end{equation}
To linearize the system (\ref{lan.39})-(\ref{lan.41}), we consider the
conformal transformation $dt=\left(  Ne^{3\omega x}\right)  d\tau$. In terms
of the evolution parameter $\tau$ the above system will be
\begin{equation}
\frac{dx}{d\tau}=\frac{\omega^{2}}{6}\left(  p_{y}-I_{2}\right)  ,~\frac
{dy}{d\tau}=\frac{1}{6}\left(  I_{2}+\left(  6\omega^{2}-1\right)
p_{y}\right)  ,~~\frac{d\zeta}{d\tau}=I_{1}e^{-2\omega y}. \label{lan.42}%
\end{equation}
Here we note that when $I_{1}=0$, then $\zeta=const$, in that case the action
integral (\ref{lan.00}) reduced to that of the scalar field cosmology.

\begin{figure}[ptb]
\includegraphics[height=8cm]{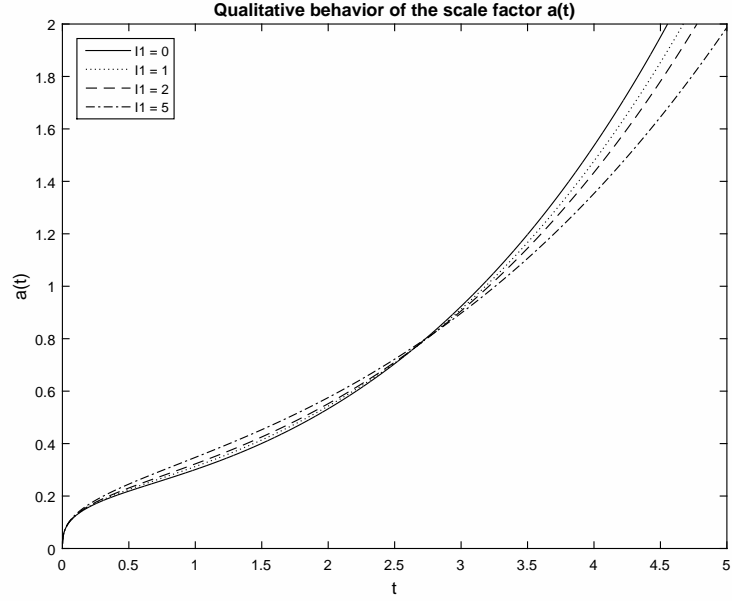}
\caption{Qualitative behavior of the scale factor $a\left(  t\right)  $, given
by the solution of the system (\ref{lan.39})-(\ref{lan.41}), for the proper
time,~$N=1$. For the plot we considered the initial condition $a\left(
0^{+}\right)  \rightarrow0$, whereas for the parameters we select,
$\omega=10^{-1}$, $I_{2}=-2$, $V_{0}=0.5$. The solid line is for $I_{1}=0$,
which corresponds to scalar field cosmology without the vector field, the dot
line is for $I_{1}=1$, the dashed line is for $I_{1}=2$, and the dash-dot line
is for $I_{1}=5.$}%
\label{scale}%
\end{figure}

\begin{figure}[ptb]
\includegraphics[height=8cm]{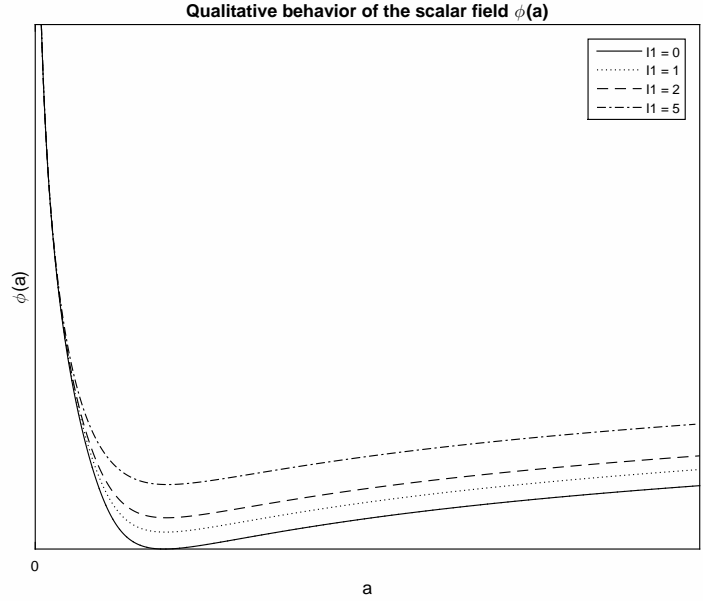}
\caption{Qualitative behavior of the scalar field $\phi\left(  a\right)  $,
given by the solution of the system (\ref{lan.39})-(\ref{lan.41}), for the
proper time,~$N=1$. The values of the free parameters are that of figure
\ref{scale}.}%
\label{scalar}%
\end{figure}

In figure \ref{scale}, we give the qualitative behavior of the scale factor
$a\left(  t\right)  $, which follows from the numerical solution of the
system, (\ref{lan.39})-(\ref{lan.41}), for the proper time, $N=1$, and for
$\omega=10^{-1}$. The plots are for different values of the constant $I_{1}$.
We observe that the existence of the vector field $\zeta$, i.e. $I_{1}\neq0$,
and for large values of the constant $I_{1}$, in the early universe the scale
factor run faster. In figure \ref{scalar} we give the qualitative behavior of
the scalar field $\phi(a)$, where we show that the behavior of the scalar
field is independent on the existence of the vector field. However what it
changes is the minimum value of the field, $min\phi$ in which increase as the
parameter $I_{1}$ increase. 

In the simplest case where $\omega=\frac{\sqrt{6}}{6}$ we obtain the
closed-form solution
\begin{equation}
x\left(  \tau\right)  =\frac{\sqrt{6}V_{0}}{36I_{2}^{2}}x_{0}e^{-\frac
{\sqrt{6}}{6}I_{2}\tau}+\frac{\sqrt{6}}{24I_{2}^{2}}\left(  x_{0}%
e^{-\frac{\sqrt{6}}{6}I_{2}\tau}\right)  ^{\frac{1}{3}}-\frac{I_{2}}{72}\tau,
\label{lan.43}%
\end{equation}%
\begin{equation}
y\left(  \tau\right)  =\frac{I_{2}}{6}\tau-\frac{\sqrt{6}}{6}\ln x_{0}%
,~\zeta\left(  \tau\right)  =-\frac{3\sqrt{6}I_{1}}{I_{0}}\left(
x_{0}e^{-\frac{\sqrt{6}}{6}I_{2}\tau}\right)  ^{\frac{1}{3}}+\zeta_{0}.
\label{lan.44}%
\end{equation}
Finally, by recalling the scale factor $a\left(  \tau\right)  =\exp\left(
\omega x\left(  \tau\right)  \right)  ,$ we find the following form for the
line element of the FRW space-time
\begin{equation}
ds^{2}=-\exp\left(  6\omega x\left(  \tau\right)  \right)  d\tau^{2}%
+\exp\left(  2\omega x\left(  \tau\right)  \right)  \left(  \delta_{ij}%
dx^{i}dx^{j}\right)  . \label{lan.45}%
\end{equation}
where $x\left(  \tau\right)  $ is given by (\ref{lan.43}).

\section{Summary}

\label{conclusion} In this paper, we studied a scalar-vector field model of
cosmology in a Lie symmetry point of view. In the presented action, in
addition to a minimally coupling between the scalar field and gravity, there
is also a coupling between the scalar and the kinetic energy term of the
vector field. For the background geometry, we have considered a flat FRW
metric and then set up the phase space by taking the scale factor $a$, scalar
field $\phi$ and the vector field $\zeta$ as the independent dynamical
variables. The Lagrangian of the model in the configuration space spanned by
$\{a,\phi,\zeta\}$ is so constructed that its variation with respect to these
dynamical variables yields the Einstein field equations. Therefore, the
Lagrangian of the cosmological model forms a three dimensional Hamiltonian
system with two unknown functions, one of which, $f\left(  \phi\right)  $,
denotes the interaction between the scalar and the vector fields while
another, $V\left(  \phi\right)  $, is the potential function of the scalar
field. To determine these unknown functions we have used a geometric criterion
known as the Lie symmetry method according to which a PDE (in our case the WDW
equation) will be invariant under the act of the generators of a Lie algebra.
Since the Lie symmetries of the WDW equation are related with the conformal
algebra of the corresponding minisuperspace, we showed that the function
$f(\phi)$ is of the form of an exponential function if the minisuperspace is
conformally flat, this also is the only case in which the WdW equation can
admit Lie point symmetries. Also, our symmetry considerations result an
exponential potential function.

This approach is also a powerful tool in finding the classical solutions to a
given Lagrangian, including the one presented above. In this approach, one is
concerned with finding the cyclic variables related to conserved quantities
and consequently reducing the classical dynamics of the system to a manageable
one. Indeed, the existence of Lie symmetry means that phase flux is conserved
along a vector field $X$ and thus a constant of motion exists.

In this set-up we studied the closed-form solutions of the WDW and the
classical field equations for a scalar-vector field cosmological model. To do
this, we applied the Lie invariants to find the invariant solutions of the WDW
equation. As we mentioned above, the importance of the application of Lie
symmetries is that they can be used in order to construct conservation laws
(Noether symmetries) for the classical field equation. Hence, we were able to
study the integrability of the field equations and to find classical
closed-form solutions. This method is related with the application of
Noether's theorem in cosmological models; however, as we have showed, it is a
more general selection rule, since, we are able to relate the classical exact
solution with the invariant form of the wave function of the universe.

\begin{acknowledgments}
AP acknowledges Prof. PGL Leach, Sivie Govinder, as also DUT for the
hospitality provided and the UKNZ of South Africa for financial support while
part of this work carried out during his visits in South Africa. The research
of AP was supported by FONDECYT postdoctoral grant no. 3160121.
\end{acknowledgments}

\end{document}